\pdfoutput=1
\documentclass[conference]{IEEEtran}
\IEEEoverridecommandlockouts
\usepackage{amsmath,amssymb,amsfonts}
\usepackage{algorithmic}
\usepackage{graphicx}
\usepackage{textcomp}
\usepackage{xcolor}
\def\BibTeX{{\rm B\kern-.05em{\sc i\kern-.025em b}\kern-.08em
    T\kern-.1667em\lower.7ex\hbox{E}\kern-.125emX}}

\usepackage[hyphens]{xurl}
\usepackage{hyperref} 

\usepackage{tcolorbox}        

\usepackage[numbers]{natbib}
\begin{document}

\title{Social Media Harms as a \textit{Trilemma}: Asymmetry, Algorithms, and Audacious Design Choices}

% \\

\author{\IEEEauthorblockN{Marc Cheong}
\IEEEauthorblockA{\textit{School of Computing and Information Systems \& Centre for AI and Digital Ethics} \\
\textit{The University of Melbourne}\\
Parkville, VIC \\
marc.cheong [at] unimelb.edu.au} 
}

\maketitle

\begin{tcolorbox}[width=3.5in,
                  %%enhanced,
                  %%frame hidden,
                halign=flush center,
                boxsep=0pt,
                  left=0pt,
                  right=0pt,
                  top=2pt]%%
This version of the manuscript is the original author-submitted version to IEEE ISTAS 2022 and not the final published conference paper. 

Please cite final  version when available on IEEE Xplore.
\end{tcolorbox}

\begin{abstract}
Social media has expanded in its use, and reach, since the inception of early social networks in the early 2000s. Increasingly, users turn to social media for keeping up to date with current affairs and information. However, social media is increasingly used to promote disinformation and cause harm. In this contribution, we argue that as information (eco)systems, social media sites are vulnerable from three aspects, each corresponding to the classical 3-tier architecture in information systems: asymmetric networks (data tier); algorithms powering the supposed personalisation for the user experience (application tier); and adverse or audacious design of the user experience and overall information ecosystem (presentation tier) – which can be summarized as the 3 A’s. Thus, the open question remains: how can we ‘fix’ social media? We will unpack suggestions from various allied disciplines – from philosophy to data ethics to social psychology – in untangling the 3A’s above.

\end{abstract}


\section{Introduction}
\label{s:Introduction}

Social media has evolved in its use and reach since the inception of early social networks in the early 2000s. Two recent issues highlight our contemporary (over-)dependence on social media. 
 
Firstly, the increasing need for social media use to stay connected\footnote{See also \url{https://en.wikipedia.org/wiki/Impact_of_the_COVID-19_pandemic_on_social_media}} and to stave away feelings of isolation and existential alienation during the ongoing COVID-19 global pandemic \cite{Carel2020-ih,Koeze2020-rv}. Social media has been used as an alternative to face-to-face contact, in line with the decision of governments around the world to discourage face-to-face contact to stop the spread of COVID-19 circa 2020-2021. 

The second is the public reaction to Facebook’s decision to block the sharing of news in Australia based on a proposed change in legislation in early 2021, which was subsequently reversed in a \textit{volte-face} \cite{Australian_Broadcasting_Corporation2021-th,Albeck-Ripka2021-rp}. What makes this case particularly pertinent is that Facebook’s user base, news agencies, and other stakeholders reacted so strongly to this turn of events. 
Reports since the 2010s in various disciplines – from information systems and computer science, to philosophy, to media studies \cite{Kwak2010-uu,DOnfro2016-ab,Alfano2021-ih,Thi_Nguyen2020-mg,Cheong2013-nj} %%(Kwak et al. 2010; D’Onfro 2016; Alfano and Sullivan 2021; Thi Nguyen 2020; Author-Redacted 2013)
have determined that users turn to social media not only for keeping in touch (i.e. maintaining their social network), but also for keeping up to date with news, current affairs and information. The title of Kwak’s paper \cite{Kwak2010-uu} even alluded to the question of whether it is “a social network or a news media?” (\textit{sic}). Contrast this with the ethos of earlier social network sites in the 1990s and 2000s, which did not have a primary role as the purveyor of news\footnote{Per an analysis of websites and URL shared on Twitter ca. 2013 \cite{Cheong2013-nj}, news sites were rarely in the top 100 top-level domains found in Twitter posts}. 

However, social media is increasingly weaponised to promote misinformation and disinformation  \cite{Ward2018-tk,Alfano2021-ih}
%(Ward 2018; Alfano and Sullivan 2021)
as well as becoming a fertile medium to propagate misinformation during the COVID-19 pandemic \cite{Marin2020-px,World_Health_Organization2020-da}
%(Marin 2020; World Health Organization et al. 2020).
This position paper argues that social media can affect the propagation of disinformation as well as actual harms through three aspects from interrelated academic disciplines. Given that social media sites are inherently large, distributed information systems, these three aspects roughly correspond to the three tiers found in the traditional information systems’ (and systems design’s) \textit{3-tier architecture}\footnote{Also attributed to John J. Donovan - see also \url{https://professordonovan.com/teaching/academic/research-academic}} \cite{Eckerson1995-yq,ScienceDirect2021-de,Siwicki1997-nf}.
%(Eckerson 1995; Siwicki 1997; ScienceDirect 2021). 

The first is from applied (social) network theory and philosophy: the inherent asymmetry of social networks can lead to a dangerous inequality of access to information, and can amplify fear and distrust \cite{Cheong2019-aap}. The second is from AI ethics\footnote{This also includes AI fairness/accountability/and transparency \cite{Diakopoulos_undated-gb}, legal studies \cite{Clifford2018-eq}, philosophy \cite{Venkatasubramanian2020-kz}, and human-computer interaction \cite{Binns2018-ux}, amongst others — which all seek to untangle the societal and human impacts of implemented algorithmic systems.} and philosophy of technology, critiquing the fact that ‘algorithms’ powering customised feeds and user engagement on social media do not ‘optimise’ for users well-being, but instead are detrimental to autonomy and epistemic well-being \cite{Alfano2020-jr,Burr2018-ig,Brincker2021-rz,Alfano2018-fc},
%%(Alfano et al. 2020; Burr, Cristianini, and Ladyman 2018; Brincker 2021; CoauthorA-Redacted, CoauthorC-Redacted, and Author-Redacted 2018)
as well as raising potential health issues \cite{Wells2021-la}. The final aspect is on tech companies' user-facing design choices, which can be termed \textit{adverse} or \textit{audacious design}. This is an umbrella term encompassing poor design choices, on two distinct levels. This ranges from the user experience: e.g. promoting frequent checking of phone app notifications \cite{Brooks2017-pu,Solon2017-ec,Jennings2006-xa};
%(Brooks 2017; Solon 2017; Jennings and Tabatabaeian forthcoming)
 to the systemic issues plaguing social media ecosystems \textit{en masse} such as poor governance structures \cite{BBC_News2021-pt}.

Briefly, the \textit{structural} mapping between social media sites, the 3-tier architecture, and relevant aspects of study concerning online disinformation can be illustrated as per Table \ref{tbl:tier}.

\begin{table}[htbp]
\caption{Social media, when seen via the 3-tier architecture, and its corresponding aspects which promote disinformation}
\begin{center}
\begin{tabular}{|p{2.5cm}|p{2.5cm}|p{2.5cm}|}
\hline
\textbf{The 3-tier architecture’s tiers} \cite{Eckerson1995-yq,Siwicki1997-nf,ScienceDirect2021-de} & \textbf{Equivalent components for a social media site} &
\textbf{Issues contributing to disinformation} \\ \hline
    Presentation: user interface & User interface 
(e.g., user experience across devices) & Adverse or Audacious design 
(e.g., how easy it is for a user to retweet fake news; how social media giants fail to address its systemic issues) \\ \hline
Application: logic rules	& Algorithms (e.g., search, personalization, feeds)	& Algorithmic issues (e.g., optimizing for attention instead of well-being) \\ \hline
Data: underlying data structures and stores	& Social network graph 
(e.g., connections)	& Asymmetric networks 
(e.g., celebrities can have many followers that react at speed and scale) \\ \hline
\end{tabular}
\end{center}
\label{tbl:tier}
\end{table}

It is important to note that for social media sites, these three aspects raised in Table \ref{tbl:tier} are not mutually exclusive; rather, they are interdependent. Each of the sections in this paper deal with only one key issue in each of these aspects, though there is no denying that there might be overlap in the discussions. For example, one cannot clearly delineate, say, the filter bubble phenomenon \cite{Pariser2011-xe} into a purely algorithmic issue (the application tier – logic rules used by the social media site to generate recommendations, say), nor a purely user interface/user experience issue (the presentation tier – how the social media site presents its recommendations to the end user, say). 

Finally, in wrapping up, this position paper discusses how we can restore trust (if it is at all possible) on social media. Each of the three tiers will again be scrutinised in an interdisciplinary lens, with possible interventions from different stakeholders: from users, to technology companies, to regulators.

%%%%%%%%%%%%%%%%%%%%%%%%%%%%%%%%%%%%%%%%%%%%%%%%%%%%%%%%%%%%%%%%%%%%%%%%

\section{Social Network Theory: The Asymmetry of Connection}
\label{s:Social}

The first argument on a major design feature of social media is that social media is increasingly becoming more asymmetric. At the 3-tier model, this lies at the very foundation: the data itself, and how it is stored and interpreted by the overall social network system.

An operative definition and background for context \cite{Cheong2019-aap} follows: early social media sites such as MySpace and the ‘friends’ feature of Facebook are symmetric, in that connections between, say, a person (denoted $X$) and another person (denoted $A$), are bilateral. $X$ enjoys the same reciprocal benefits from ‘friending’ $A$ on Facebook, and vice versa. In graph notation, it can succinctly be described as $X \longleftrightarrow A$.

Over time, on newer social media sites, and due to changes in functionality on existing social media sites (including the introduction of Facebook Fan Pages), the asymmetric mode of connection \cite{Cheong2019-aap} came to the forefront. Let’s say one admires a celebrity (say $C$) and would like to connect with them online. $X$’s ‘following’ of $C$ — on Facebook’s Pages, Instagram, Twitter — is asymmetric, in the sense that the reciprocal value is significantly lower (if not zero). In other words, celebrity $C$ need not be interested in $X$’s status updates nor engage her in any way; it’s only the unilateral action of $X$ in the relationship ($X \longrightarrow C$) which matters \cite{Cheong2019-aap}.

Thus, effects of the unilateral nature of this asymmetric connection include, firstly, the ability of a popular user $C$ in the network to attract new followers etc; in other words, the ‘asymmetry of attachment’, or ‘rich get richer’ effect \cite{Cheong2019-aap,Barabasi1999-nm,Green2014-kk}
%(Barabási and Albert 1999; Green 2014; Author-Redacted 2019).
However, in practise, this leads to what is termed the ‘asymmetry of influence’ effect proposed in \cite{Cheong2019-aap}.  ‘Like’s, replies to, and retweets for celebrities are seen as ‘currency’ which leads to the ability to command inequitable attention to spur both on- and offline action. Online ‘pile ons’ \cite{White2020-hw,Coscarelli2018-rr}%(White 2020; Coscarelli 2018)
, toxic comments, and lack of accountability in ‘mob action’ are some negative examples. The compression of valuable communication and social acknowledgement into quantifiable metrics, or gamification as Thi Nguyen \cite{Thi_Nguyen2021-mu} puts it, “increases our motivation by \textit{changing the nature of the activity}… invit[ing] us to shift our values along its pre-fabricated lines... [when users] start to \textit{chase higher Likes and Retweets and Follower counts}” (\textit{emphases mine}).

The question is: what issues can be raised in relation to the changing landscape of information that is one of the focal points of this paper? Per Section \ref{s:Introduction}, social media has been increasingly treated as a source of news: the asymmetries discussed here mean that newsmakers and agenda setters — however their political leaning — has the ability to influence the masses who subscribe to them, via their social media following. This might work for good, e.g. for celebrities to advance a particular noble cause; an example is Lady Gaga\footnote{See \url{https://www.bbc.com/news/entertainment-arts-52199537}}  \cite{Archer_undated-nn}.

Conversely, it might also be dangerous, as in the case of malevolent action such as election-based and health misinformation -- such as in the case of Donald Trump --  if advocated by celebrities \cite{Twitter_Inc2021-cb,Ward2018-tk,Jetter2020-fh}.
%(Twitter Inc. 2021; Ward 2018; Jetter, Lewandowsky, and Ecker 2020).
The asymmetries discussed above are also compounded by the fact that algorithms determine what contents users want to see, as part of their user experience.

%%%%%%%%%%%%%%%%%%%%%%%%%%%%%%%%%%%%%%%%%%%%%%%%%%%%%%%%%%%%%%%%%%%%%

\section{Algorithms of “Math Destruction”}
\label{s:Algorithms}

Following on from an exposition of asymmetric network properties, this section will cover how algorithms play a role in compounding social media mis- or disinformation, as a `\textit{weapon of math destruction}'\footnote{To paraphrase Cathy O’Neil’s book, algorithms are capable of “math destruction” \cite{ONeil2016-xn}}. A succinct introduction is as follows: algorithms are merely collections of complex rules, or mathematical programs that work on vast amounts of user data, deployed at tech companies. The popular conception of ‘algorithm’ expands its scope to the application/deployment of such programs and its overall impacts on society \cite{Noble2018-il}.% (Noble 2018). 

The asymmetries mentioned in Section \ref{s:Social} are compounded by the use of strategic algorithms by social media companies to curate what they ‘think’ we want to see. As alluded in the Introduction (Section \ref{s:Introduction}), it is hard to cleanly delineate between the algorithms and the design of the system. Hence, to clarify, this section will only focus on issues closely related to the algorithm – in other words, the rules and application logic powering social media sites’ functionality (search, personalisation, recommendation, amongst others). The adverse design choices, combined with overall ‘top-down’ design and how a user perceives and interacts with the social media system, will be covered in Section \ref{s:Audacious}.

An infamous example of issues with ‘the algorithm’ is the Facebook News Feed algorithm, which started out being a simple collection that showed users “personal notifications” \cite{DOnfro2016-ab} with a chronological focus 
\cite{Murphy2013-gi}, not dissimilar to how an information systems presents, say, database query results. However, the ongoing evolution of Facebook's News Feed, which is one of the earliest attempts to personalise and curate feeds for its users, has opened a Pandora’s Box. 

Starting from endeavours to “predict[...] which stories you would be most interested in seeing” \cite{Murphy2013-gi} has given it “...an unprecedented amount[...] of power over people's digital lives” \cite{DOnfro2016-ab}. The 2006 introduction of the News Feed met with user backlash \cite{DOnfro2016-ab,Murphy2013-gi}, prescient of the various issues concerning algorithmic personalisation we are still facing. Over time, this algorithm is no longer predictable nor explainable from the outside looking in; it is a proprietary system trained upon modern techniques such as deep machine learning in order to optimise for user engagement. In fact, a popular conception about the ‘algorithm’ is that it’s something unpredictable but yet crucial to ‘work with’ in drawing attention to one’s social media presence, especially when there is monetisation at stake \cite{Byrne2019-yw,Rodriguez2019-fh}.

In other words: such algorithms do not optimise for user knowledge or epistemic well-being \cite{Millar2019-ji,Alfano2021-ih} such as presenting a diverse range of trustworthy credible sources and debunking falsehoods \cite{Samantray2019-aa}. Social media optimises the provision of ‘dopamine\footnote{A clear discussion on the effects of brain neurochemistry with respect to technology use is at \url{https://sitn.hms.harvard.edu/flash/2018/dopamine-smartphones-battle-time/}} hits’ \cite{Brooks2017-pu} to encourage ongoing use. Or, quoting Sean Parker of Facebook, to “...consume as much of your time and conscious attention as possible” \cite{Solon2017-ec}. Ostensibly this engagement is currency: with improved monetisation opportunities for various stakeholders \cite{Martin2011-bv}.  

This poses a challenge to us as consumers: in what way can the algorithms lead to actual harms? Documented effects on one’s health include, for one, addiction \cite{Brooks2017-pu,Solon2017-ec}, which may lead to poor health outcomes \cite{Tarsha2016-sn}. This is, sadly, an ongoing concern. With regards to user autonomy, these algorithms – however subtle – introduce the prospect of normalising behavioural change (raised earlier in Section \ref{s:Social}) in order to conform to their ‘optimised’ modus operandi, in the interests of the technology companies themselves. In fact, leaked Facebook research (Section \ref{s:Slaying}) claims that “that 1 in 8 of its users report engaging in compulsive use of social media that impacts their sleep”, affecting “about 360 million users” \cite{Wells2021-oh}, an alarming statistic indeed.

Another point pertaining to this position paper is harm to our intellectual capability as ‘knowers’. Two popular examples will be discussed firstly: firstly, epistemic harms in the form of filter bubbles \cite{Pariser2011-xe} and more insidiously, echo chambers \cite{Thi_Nguyen2020-mg}. Simply put, ‘filter bubbles’, attributed to Pariser \cite{Pariser2011-xe}, are when we keep viewing what the machine ‘thinks’ (or includes in its filter) we want to see while ‘filtering out’ other views\footnote{ Thi Nguyen \cite{Thi_Nguyen2020-mg} provides an excellent definition of the various phenomena in the introductory section of their paper.}: this feedback loop continues ad infinitum when we act upon its recommendations. Echo chambers are more insidious in the fact that those caught within “systematically distrust all outside sources” \cite{Thi_Nguyen2020-mg}, which happens in, say, forums where participants are engrossed in online conspiracy theories. 

Another way this harm manifests itself is illustrated in our overdependence on social media as a news source — per our argument on encouraging ongoing use — which makes it hard for us to wean ourselves off it \cite{Dick2021-hp}. This is clearly illustrated by the public reaction to Facebook’s block on news sources in Australia, as alluded in Section \ref{s:Introduction}. The huge backlash from consumers and other stakeholders \cite{Australian_Broadcasting_Corporation2021-th} reveals the extent to which users overly depend\footnote{It remains to see how Australian social media consumers evolve their news consumption habits, for better or for worse, as an aftereffect of Facebook’s initial ban and subsequent ‘about face’.} on social media as a form of news. More importantly, this block,  if it persisted, can introduce a news vacuum \cite{Dick2021-hp} which can – paradoxically – lead to the proliferation of misinformation and fake news \cite{Cave2021-bn}.

%%%%%%%%%%%%%%%%%%%%%%%%%%%%%%%%%%%%%%%%%%%%%%%%%%%%%%%%%%%%%%%%%%%%%%%%

\section{Audacious, or Adverse, Design?}
\label{s:Audacious}

The last aspect this position paper covers is the design of the user experience of social media apps and ecosystems. This is not limited only to, say, the user-facing design of these apps (say, the graphical interface to the retweeting functionality of Twitter), but rather encompasses the ‘big picture’ perspective of a users’ ongoing interaction with the social media platform (to extend the example: the ongoing engagement and use of the Twitter app, from ‘sharing’ functionality, to overall governance of sharing behaviour of the user base). 

This section is divided into two – firstly, what we term audacious design choices, stemming from the dictionary definitions of ‘bold’ choices which exhibit a ‘lack of respect’ for the user \footnote{  Per the Oxford Dictionary of English, via MacOS’s Dictionary app.}. These are primarily sourced from the recent \textit{Facebook Files}\footnote{  A detailed general overview is at \url{https://en.wikipedia.org/wiki/Facebook_Files}} exposé from ex-Facebook whistle-blower Frances Haugen in The Wall Street Journal \cite{The_Wall_Street_Journal2021-be}. The second subsection deals with design choices, or countermeasures\footnote{  This term is adopted in the spirit of Marin \cite{Marin2020-px}, who provides an excellent discussion on the topic.}, deployed by social media companies to curb misinformation. Unfortunately, we argue that these steps are (sadly) reactionary, rather than proactive, in the face of stakeholder pressures.

\subsection{Audaciousness in Design}\label{AA}
As seen in the Facebook Files exposé, many problems abound in the overall design of not just the (user-facing) aspect of social media sites, but the design of human oversight and accountability structures, which shape the user experience. These can easily be generalisable to any generic social media site.

For instance, “politicians and high-profile Facebook users had different rules governing what content they could post” \cite{BBC_News2021-pt}, contradicting their public persona of having “standards of behavio[u]r apply to everyone, no matter their status or fame” \cite{Horwitz2021-tx}. 

Facebook is also accused of having error-prone systems “with only minimal success in removing hate speech, violent images and other problem content” \cite{Seetharaman2021-au}. In terms of user content, it allowed “plagiarized and recycled content to flourish on its platform despite having policies against it” \cite{Hagey2021-lb}; was accused of “show[ing] people positive stories about the social network… [by] pushing pro-Facebook news items — some of them written by the company… [to] would improve its image” \cite{Mac2021-lj}; while simultaneously accused of exposing users to incendiary and controversial content in the name of engagement \cite{Rai2021-cf}. 

\subsection{Remedying Adverse Design}

When it comes to responding to public pressure when faced with a deluge of misinformation (primarily during the COVID-19 pandemic), social media companies are quick to claim to have addressed these issues with design countermeasures, mostly in their interface design. A few such countermeasures are highlighted here. First, on a user-level, is the \textit{vouching} of bona fide users on such platforms, and \textit{disavowing} of users who sow misinformation \cite{Quintana_undated-sc}, which may be conducive to building trust. 

Consider the ‘vouching’ approach using a ‘verified’ user badge on Facebook and Twitter, only given to users vetted by social media companies\footnote{ For the full details of the process, see \url{https://help.twitter.com/en/managing-your-account/twitter-verified-accounts} (Twitter) and \url{https://www.facebook.com/help/1288173394636262?helpref=faq_content} (Facebook).}. The bad news is that this system isn’t foolproof: consider the 2020 Twitter hack which has seen ‘verified’ accounts of the “rich and famous…” being commandeered in a cryptocurrency scam \cite{Leins2020-bn}.

Social media companies also disavow users via suspensions and bans. An infamous example is Twitter’s “permanent suspension” of Donald Trump “due to the risk of further incitement of violence” \cite{Twitter_Inc2021-cb}. However, critiques are made of social media companies’ inconsistencies (Section \ref{ss:Asymmetry}), with many claiming they are “not doing enough... [and]... slow to act” \cite{Stevenson2018-dz}.

The second form of countermeasure is on a ‘message’ level, specifically fact-checking. In line with the COVID-19 pandemic, social media companies’ “efforts for fact-checking were accelerated to an impressive extent” \cite{Marin2020-px}. Efforts to fact-check political information were also noticed by journalists, with claims that “Facebook... [has] been more active in its steps to prevent the platform from unduly impacting the 2020 election” \cite{Mantas2020-cz}. 
These measures are praiseworthy, but reflect two concerns: social media companies have to willingly and fully adopt the role of gatekeeper (not just paying lip service, reacting to stakeholder pressure); and the inherent technological limitations of fact-checking \textit{everything} \cite{Croce2021-vd,Alam2020-li}. For the former, the hesitance of social media companies to fully assume these responsibilities \cite{DOnfro2016-ab} is a stumbling block\footnote{This again ties in with the Australian Facebook news ban.} .

\section{‘Slaying the Hydra’: Countermeasures, or Counterintuitive solutions?}
\label{s:Slaying}

What, then, are solutions that we can expect from social media sites, given the three interdependent facets of social media – the asymmetric network; dangerous algorithms; and adverse/audacious overarching design choices – analogous to the many-headed \textit{Hydra} from ancient mythology? Here, several possible solutions are presented as a mere starting point for addressing this trilemma.

\subsection{Asymmetry}
\label{ss:Asymmetry}

As seen in Section \ref{s:Social}, the asymmetric mode of social networking today is problematic. A potential response is to deprioritise the asymmetry of influence for popular users. The move by Instagram to remove the count of ‘likes’ on posts is one step in the right direction, to help steer Instagram away from becoming a popularity contest, and reduce cyberbullying \cite{BBC_News2019-ja}. A similar move is made on YouTube, where the `dislike' counts on a video are hidden to the public\footnote{See YouTube's explanation at \url{https://blog.youtube/news-and-events/update-to-youtube/}}.

The motivations behind them, however, are in doubt. Is it simply trying to generate more ad revenue instead, as criticised by some \cite{Rodriguez2019-fh} in the Instagram case?  This is only a drop in the ocean: in the recent Facebook Files, Facebook knows that Instagram has “a significant teen mental-health issue that [it]… plays down in public” \cite{Wells2021-la}. Similar arguments exist for YouTube's decision -- is it just paying lip service to mental health \cite{Suciu2021-dg} by removing the `dislikes'?  

\subsection{Algorithms}
Also, as seen in Section \ref{s:Algorithms}, Facebook was one of the pioneers of personalised feeds. But what if the user base was given the capability of fully fine tuning and optimising their own feeds for their own wellbeing; prioritising news from credible sources without influence of advertisements of sponsored posts; or having algorithms up for audit by epistemologists and data scientists? These would likely be against social media companies’\textit{ raison d’etre} of creating value and revenue! 

\subsection{Adverse/Audacious Design}
The design of social media platforms – both the user experience on their apps, as well as overarching systems and structures of governance – is a challenging aspect to tackle. For the former, suggestions range from “structural approaches” to change the “architecture of online environments” to counter fake news \cite{Croce2021-lk,Croce2021-vd}; to incorporating active steps in the user experience to educate or ‘inoculate’ users on how to avoid fake news and micro-targeting on social media \cite{Lewandowsky2021-es,Cook2017-zl}; to introducing psychological science-guided interventions to user experiences \cite{Kozyreva2020-wi}.

For the latter, ambitious proposals include treating social media platforms as monopolies which can be “nationalized, highly regulated, or broken up” \cite{Alfano2021-ih}; or the development of an open standard for decentralised social media which supports competition, similar to how modern email services are structured, where no single company exerts executive control.

\section{Conclusion: Rebuilding Trust or Back to Basics?}
This position paper has covered how social media has parallels with the 3-tiered architecture for systems design. The three factors – the large-scale asymmetric structure of social media, algorithmic factors and non-optimisation for well-being, and social media’s adverse/audacious design choices – contribute to the spread of disinformation, erosion of trust, and a myriad of other harms to users.

In summary, from observing these three aspects, technology companies’ design choices for each of the three tiers have profound emergent effects on our agency as consumers of information and autonomous consumers of social media. The ethos of optimising for engagement, which leads to increased profits, is at diametric opposites with optimising for user well-being and knowledge.

For users, are we truly at the mercy of social media companies? By gaining even the most basic awareness on the issues plaguing social media (such as those raised in the Facebook Files), users can start empowering themselves to minimise the harm – from epistemic concerns, to actual health concerns – resulting from social media consumption.

We can start by flexing our epistemic muscles, and our autonomy: we have to exercise critical thinking in the process of selecting our news sources, determine credibility of viewpoints, uncovering media biases, and knowing when to go on a social media diet, for starters.

\section*{Acknowledgment}
This paper is based on unpublished talks the author presented in 2021: (1) at the EU-funded PERITIA ``Trust in Expertise in a Changing Media Landscape'' conference (March 2021); and (2) the CIS Seminar series at the School of Computing and Information Systems, the University of Melbourne (July 2021).

The author acknowledges the PERITIA Project and all fellow conference panelists who have provided helpful suggestions and ideas, including Ty Branch, Hossein Derakhshan, Michael Geers and team (especially Stephan Lewandowsky and Philipp Lorenz-Spreen). At the University of Melbourne, the author thanks Reeva Lederman and Ofir Turel for their suggestions from the CIS Seminar.
 
\bibliography{paperpile2}
\end{document}